\documentclass[a4paper]{article}

\usepackage{INTERSPEECH2020}

\title{Memory Controlled Sequential Self Attention for Sound Recognition}
\name{Arjun Pankajakshan$^{1}\thanks{AP is supported by a QMUL Principal's studentship. EB is supported by a Turing Fellowship. This work has received funding from the EU’s Horizon 2020 research and innovation programme under the Marie Skłodowska-Curie grant agreement No. 765068.}$,
       Helen L. Bear$^{1}$, Vinod Subramanian$^{1}$,
       Emmanouil Benetos$^{1,2}$
       }
\address{$^1$ School of EECS, Queen Mary University of London, UK\ \ \ \  $^2$ The Alan Turing Institute, UK}
\email{\{a.pankajakshan, h.bear, v.subramanian, emmanouil.benetos\}@qmul.ac.uk}

\begin{document}

\maketitle
\begin{abstract}
In this paper we investigate the importance of the extent of memory in sequential self attention for sound recognition. We propose to use a memory controlled sequential self attention mechanism on top of a convolutional recurrent neural network (CRNN) model for polyphonic sound event detection (SED). Experiments on the URBAN-SED dataset demonstrate the impact of the extent of memory on sound recognition performance with the self attention induced SED model. We extend the proposed idea with a multi-head self attention mechanism where each attention head processes the audio embedding with explicit attention width values. The proposed use of memory controlled sequential self attention offers a way to induce relations among frames of sound event tokens. We show that our memory controlled self attention model achieves an event based $F$-score of $33.92\%$ on the URBAN-SED dataset, outperforming the $F$-score of $20.10\%$ reported by the model without self attention.
\end{abstract}


\section{Introduction}
Sound event detection (SED) \cite{virtanen2018computational} is the task of automatic transcription of sound event tags with onset and offset positions from audio sequences. The essential architectural block of a deep neural network based SED model is the convolutional recurrent neural network (CRNN) \cite{cakir2017convolutional}. The convolutional layers extract frame level features that are invariant to local spectral and temporal variations. The frame level features are sequentially processed by the recurrent layers to model relations among frames within the input sound sequence. However standard recurrent neural networks (RNNs) have two drawbacks. Firstly, in RNNs the recursive state update is performed in a first order Markov manner, which lacks an adaptive memory control mechanism. To explain this, long term memory is required when there exist relations among sound events at distant positions in long sequences. On the other hand, to process shorter sequences and in the case when relations among sound events are not certain, long term memory is not needed. The frame level audio features given to the recurrent layers are highly correlated over time, consequently the recursive state updates without adaptive memory control may result in an improper summarisation of the sound event sequence. Another undesired property of RNNs is the lack of a mechanism for modeling relations between audio frames in a sound event sequence. In sound recognition this omission, added with the fact that sound event sequences lack inherent structure, is a big limitation in sound event sequence modeling. 

Attention mechanisms address these RNN limitations and have become an intrinsic part of neural network models in various tasks such as neural machine translation (NMT) \cite{bahdanau2014neural,luong2015effective,vaswani2017attention}, machine reading \cite{cheng2016long}, image captioning \cite{xu2015show}, image synthesis \cite{zhang2018self}, and speech recognition \cite{chorowski2015attention,bahdanau2016end}. \emph{Self attention} \cite{vaswani2017attention,cheng2016long} is an attention mechanism that models the relations within a single sequence to compute a better summarisation of the sequence. %In this work we investigate the importance of attention width %address two questions; 1) What is the significance of recurrence relations among sound events in sound event sequences? 2) What is thein sequential self attention for better summarisation of the sound event sequence%
When recurrence relations are persistent throughout a sequence (regardless of the dimension of the feature embedding i.e., the feature embeddings are either in the form of event-level tokens or frame-level tokens), then any choice of attention width induces relations within the sound event sequence. However, recurrence relations are uncertain in sound event sequences, so we assume that long-term memory is not needed in sound event sequence modeling. Therefore we propose \emph{memory controlled self attention} to learn better latent representations of sound event sequences.

Sequential attention mechanisms \cite{bahdanau2014neural} jointly translate and align words using \emph{global} or \emph{soft} attention. This is when all the encoder hidden states with different attention weights are used to predict the decoder output at each timestep. Luong et al. \cite{luong2015effective} proposed \emph{local} attention that selectively focuses on a small context window of the encoder to predict decoder outputs. In the context of self attention, \emph{local} attention and our memory controlled attention are the same. \emph{Sequential self attention} \cite{cheng2016long} has successfully been applied to machine reading, using a memory network with non-Markov recursive state updates. The attention function is more generally described as mapping a \emph{query} and set of \emph{key-value} pairs to an output \cite{vaswani2017attention}. The set of keys and values define the extent of the memory used for attention. To the best of our knowledge, none of these works have analysed the extent of memory on attention performance for the respective tasks. We assume that the extent of memory (attention width) is not influential in the context of speech and text data because of the persistent relations between word tokens in these data sequences added with the auto-regressive modeling power of these models. 

Attention mechanisms have been used for sound recognition; for example in temporal attention for audio tagging  \cite{xu2017attention,yu2018multi}, attention and localization are used to quantify sound events at each audio frame. Kong et al. \cite{kong2018audio,kong2019weakly} proposed an attention model for multiple instance learning (MIL) applied to audio classification. In SED, Wang et al. \cite{wang2018self} applied self attention mechanisms based on \emph{transformer attention} \cite{vaswani2017attention}. Again, the authors have not investigated the impact of the extent of memory (key-value selection) on attention performance. Interestingly, their work shows that overall detection does not improve with the self attention mechanism. But also, their self attention implementation improved the detection performance for some long duration sound events. This indicates a need for memory controlled self attention in sound recognition. 

In this paper, we evaluate the potential of memory controlled sequential self attention for sound event detection; we also propose a methodology to quantify a range of attention width values to summarise each audio frame embedding using \emph{multi head self attention}. To the best of our knowledge, there has not been work exploring the use of sequential self attention mechanisms for SED. The rest of this paper provides a description of memory controlled self attention, our multi head attention proposal, followed by the experimental details, results and discussion. 

\section{Motivation}
By comparing various aspects of sequence modeling of audio signals with sequence modeling in natural language processing (NLP), in this section we aim to show that memory controlled self attention is an appropriate choice for sound event sequence modeling.

\begin{itemize}
\item Similar to speech and music signals, sound event sequences belong to the class of structured sequence data; however recurrence relations are uncertain in sound event sequences. This means, it is not prudent to assert relations between consecutive sound events in these sequences. However, there exist temporal relations between audio frames within sound events.
\item In speech and text data the relations between phonemes in a word and the relations between consecutive words in a sentence are assured (i.e., recurrence relations are persistent in text data regardless of the dimension of the feature embedding). Hence self attention with any memory width is unambiguous for general NLP applications. The language structure and the semantic relations in text data support this behaviour.
\item In speech and text processing, self attention is generally applied on word level embeddings \cite{wu2018phrase, cheng2016long, bahdanau2014neural, vaswani2017attention}. Contrarily in sound recognition, self attention is applied to frame level embeddings. We claim that the lack of higher level event-based embeddings is the most important constraint in sound event sequence modeling. The frame level features in sound event sequences are highly correlated over time. Thus SED models without adaptive memory controlled self attention may overfit to pseudo relations based on frame level similarity patterns. This reduces the effectiveness of self attention mechanisms and lessens the recognition performance and generalizability of sound recognition models.
%Considering the internal mechanism of self attention, the attention function on the current frame level token is the aggregation of similarity scores of the current token with respect to all the other tokens within the attention width. If the  self attention memory width is not set properly, it is highly likely that the current frame of the current sound event is incorrectly influenced by a far distant frame of a different sound event where the two items would lack any recurrence relations.%
\end{itemize}

\section{Memory Controlled Self Attention}

We implement memory controlled sequential self attention on top of a CRNN model for the task of sound event detection. The architectural details of the CRNN model are described in Section $4.1$. The convolutional block maps an audio input sequence of representations $\textbf{X} = (\textbf{x}_1, \textbf{x}_2, \cdots, \textbf{x}_T)$ to a sequence of feature embeddings $\textbf{Z} = (\textbf{z}_1, \textbf{z}_2, \cdots, \textbf{z}_T)$, where $T$ is the total number of audio frames. Given $\textbf{Z}$, at each time step the recurrent layer generates hidden state representations $\textbf{H} = (\textbf{h}_1, \textbf{h}_2, \cdots, \textbf{h}_T)$. In this work, we apply the proposed memory controlled self attention layer on $\textbf{H}$ to derive improved hidden state representations $\tilde{\textbf{H}} = (\tilde{\textbf{h}}_1, \tilde{\textbf{h}}_2, \cdots, \tilde{\textbf{h}}_T)$ prior to classification.

A self attention function on an input sequence is described as mapping each query vector of the sequence with a set of key-value vectors to obtain an output vector that summarises the query vector with respect to the key-value set. The output vector is the weighted sum of the value vectors, where the weight assigned to each value vector is computed by a similarity function of the query vector with the corresponding key vector. Using the general form of the self attention function without memory control, each bottleneck feature vector $\tilde{\textbf{h}}_t$ is computed as:
\begin{equation}
\begin{split}
  \tilde{\textbf{h}}_t &= \sum_{i=1}^{T}{\alpha_i^t}{\textbf{h}_{i}}; \allowbreak t \in \{1,\ldots,T\}
  \label{eq1}
\end{split}
\end{equation}
where $\alpha_i^t$ is the attention weight value computed using a similarity function as:
\begin{equation}
\begin{split}
  {\alpha_i^t} & = softmax({s_i^t}) \\
  {s_i^t} & = score(\textbf{h}_t, \textbf{h}_i); \allowbreak i,t \in \{1,\ldots,T\}\\
\end{split}
\label{eq2}
\end{equation}
\begin{equation}
\begin{split}
  score(\textbf{h}_t, \textbf{h}_i) =
    \begin{cases}
      \textbf{v\textsubscript{a}}^{\intercal}tanh(\textbf{W\textsubscript{a}}[\textbf{h}_t;\textbf{h}_i]) & \text{additive/concat}\  \cite{bahdanau2014neural}\\
      \textbf{h}_t^{\intercal}\textbf{W\textsubscript{a}}{\textbf{h}_{i}} & \text{general}\ \cite{luong2015effective}\\
      \textbf{h}_t^{\intercal}\textbf{h}_i & \text{dot}\ \cite{luong2015effective}\\
      \textbf{h}_t^{\intercal}\textbf{h}_i/ \mid \textbf{h}_t \mid \mid \textbf{h}_i \mid & \text{scaled dot}\  \cite{vaswani2017attention}
    \end{cases}       
\end{split}
\label{eq3}
\end{equation}
where $\textbf{v\textsubscript{a}}$, $\textbf{W\textsubscript{a}}$ are the weight terms of the score functions and ${\intercal}$ denotes transposition.

To explain (\ref{eq1}), the general form of self attention computes the similarity of each frame level embedding with respect to every other feature embedding in the input sequence. The similarity scores between frame level embeddings of distinct sound event tokens might be high, which results in a wrong summarisation of the input sequence. Also, as sound event tokens in audio sequences typically lack syntactic and semantic relations, long term memory is not required. However relations exist among frame level embeddings within sound event tokens, thus to effectively model these relations we propose \emph{memory controlled self attention} by constraining the self attention function in (\ref{eq1}) to a compact neighbourhood relative to each frame level embedding with $L$ being the attention width.
\begin{equation}
  \tilde{\textbf{h}}_t = \sum_{i=(t-(L/2))}^{t+(L/2)}{\alpha_i^t}{\textbf{h}_{i}}; \allowbreak t \in \{1,\ldots,T\}
  \label{eq4}
\end{equation}
In terms of \emph{query}, \emph{key}, and \emph{value} representations we have $\textbf{Q}$ = $\textbf{h}_t$, and $\textbf{K\textsubscript{L}}$ = $\textbf{V\textsubscript{L}}$ = $(\textbf{h}_{t-(L/2)}, \cdots, \textbf{h}_t, \cdots, \textbf{h}_{t+(L/2)})$. The key-value set determines the extent of self attention. Using this we update the general form of memory controlled self attention equivalent to that of (\ref{eq4}) as:
\begin{equation}
  Attention(\textbf{Q}, \textbf{K\textsubscript{L}}, \textbf{V\textsubscript{L}}) = softmax(\textbf{Q}{\textbf{K\textsubscript{L}}}^{\intercal})\textbf{V\textsubscript{L}}
  \label{eq5}
\end{equation}

We evaluate the impact of memory controlled self attention on sound recognition performance using the different score functions listed in (\ref{eq3}). In preliminary experiments, we achieved the best results using the additive score function. Therefore, the results and observations included in this paper are based on the additive score function.

A limitation of the memory controlled self attention function in (\ref{eq4}) is that it uses a fixed attention width value to summarise each frame level embedding independent of the duration of sound events. However it is better to use a small attention width value to the frames that belong to sound events that have short duration and a large attention width value to the frames associated with long duration sound events. Ideally the best memory controlled self attention design would automatically choose appropriate attention width values to summarise each frame level embedding in the input sequence. We therefore propose multi head memory controlled self attention to address this limitation.

\subsection{Multi-Head Self Attention}

As an alternative to using a fixed attention width value, we propose to apply the same attention function on each query with different key-value sets. A key-value set with the corresponding attention width value leads to a memory controlled self attention head. The implementation of the multi head memory controlled self attention function is depicted in Fig.~\ref{fig:multi-head}. The weight of each head is normalised using its corresponding attention width value. The normalised weighted sum of the attention head output values wrap up the final frame level embeddings as:
\begin{figure}[t]
  \centering
  \includegraphics[width=\linewidth,height=4.5cm,keepaspectratio]{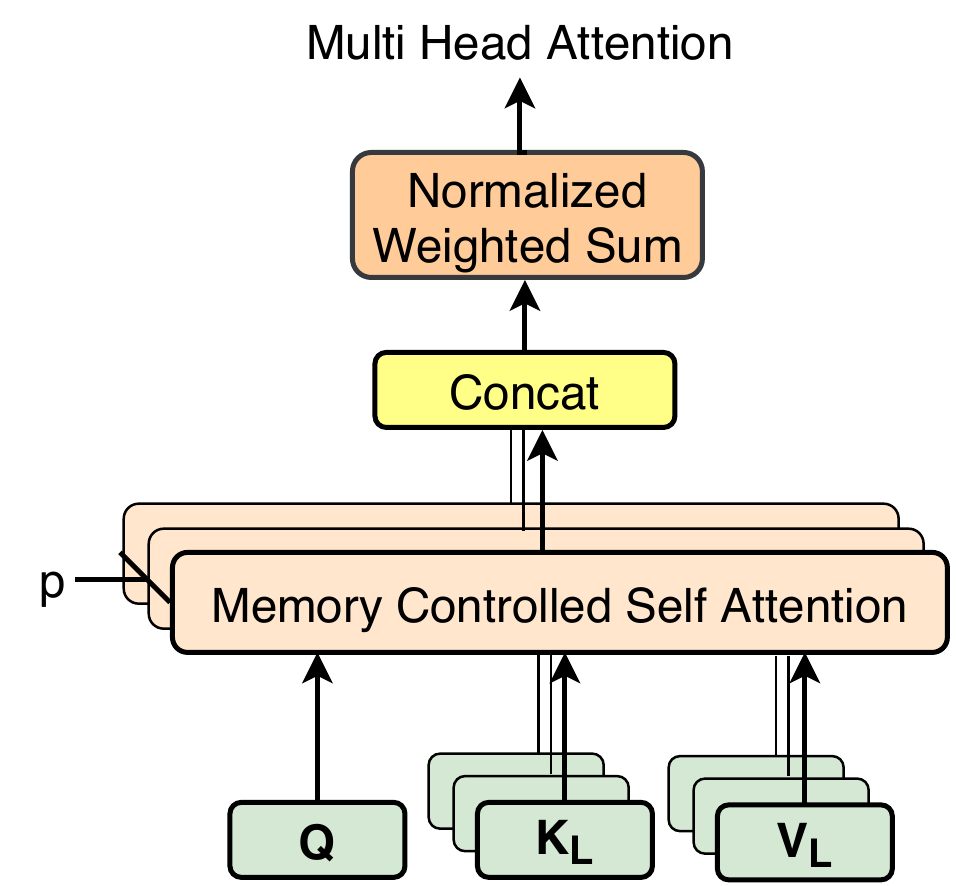}
  \caption{Block diagram of Multi Head Memory Controlled Self Attention.}
  \label{fig:multi-head}
\end{figure}
\begin{equation}
\begin{split}
  MultiHead(\textbf{Q}) &= Concat(\textbf{head}_{1}, \ldots, \textbf{head}_{p})\textbf{W\textsubscript{ah}} \\
  \textbf{head}_{j} &= Attention(\textbf{Q}, \textbf{K\textsubscript{L}}_{j}, \textbf{V\textsubscript{L}}_{j}); \allowbreak j \in \{1,\ldots,p\}\\
  \textbf{W\textsubscript{ah}} &= (w_1/L_1, \cdots, w_p/L_p)
  \label{eq6}
\end{split}
\end{equation}
where $p$ is the number of attention heads, $Concat$ denotes the concatenation of individual attention head vectors, $\textbf{W\textsubscript{ah}}$ is the normalised weight vector with $w_j$, $L_j$ respectively denoting the weight and attention width values for the $j^{\text{th}}$ head.

\noindent \textbf{Comparison to transformer multi head attention} \cite{vaswani2017attention}: Whilst our multi head self attention implementation in (\ref{eq6}) is similar to the \emph{Transformer multi head attention} in (\ref{eq7}), there are a few critical differences. To the best of our knowledge, there has not been any other work exploring multi head architectures for self attention. Firstly, each of our self attention head has a corresponding key-value set that determines the extent of self attention for that head. Hence, our multi head attention approach implements a soft optimization rule to rank individual attention heads for the best summarisation of the frame level embeddings. \emph{Transformer multi head attention} linearly projects the same key-value set with different learned weights at each attention head:
\begin{equation}
\begin{split}
  MultiHead(\textbf{Q}, \textbf{K}, \textbf{V}) &= Concat(\textbf{head}_{1}, \cdots, \textbf{head}_{p})\textbf{W\textsubscript{O}} \\
  \textbf{head}_{j} &= Attention(\textbf{Q}\textbf{W\textsubscript{Q}}_{j}, \textbf{K}\textbf{W\textsubscript{K}}_{j}, \textbf{V}\textbf{W\textsubscript{V}}_{j});\\
  & j \in \{1,\ldots,p\}\\
  Attention(\textbf{Q}, \textbf{K}, \textbf{V}) &= softmax(\textbf{Q}\textbf{K}^{\intercal}/\sqrt{d_k})\textbf{V}
  \label{eq7}
\end{split}
\end{equation}
where $\textbf{W\textsubscript{O}}$, $\textbf{W\textsubscript{Q}}_{j}$, $\textbf{W\textsubscript{K}}_{j}$, $\textbf{W\textsubscript{V}}_{j}$ are the weight matrices and $d_k$ is the dimension of the key vector. Secondly, our multi head implementation has only a single attention layer with score function weights ($\textbf{v\textsubscript{a}}$ and $\textbf{W\textsubscript{a}}$ in (\ref{eq3})) and attention head weight ($\textbf{W\textsubscript{ah}}$ in (\ref{eq6})). \emph{Transformer attention} \cite{vaswani2017attention}, on the other hand, has separate attention head layers with associated weight matrices as shown in (\ref{eq7}). Lastly, we compute attention weights using the additive score function of (\ref{eq3}), whereas in \cite{vaswani2017attention} the scaled dot product score function is used. In our multi head approach, for $L_a$ $>$ $L_b$, $Attention(\textbf{Q}, \textbf{K\textsubscript{L}}_{a}, \textbf{V\textsubscript{L}}_{a})$ is a superset function of $Attention(\textbf{Q}, \textbf{K\textsubscript{L}}_{b}, \textbf{V\textsubscript{L}}_{b})$. This may result in biased attention head weight assignment. To counteract this effect, the attention weight of each head is scaled with the corresponding attention width value.

In this work we empirically choose $p$ = $11$ attention heads in the multi head self attention layer. The first attention head employs an attention width of $L=2$ to observe the impact of immediate past and immediate future frame level embeddings to summarise the present frame. In the subsequent attention heads we serially increment the attention width value by five frames.

\section{Experimental Details}

We first analyse SED performance with a standard self attention function as in (\ref{eq1}). Then we analyse the impact of memory controlled self attention (\ref{eq4}) with different attention width values on SED performance. Lastly, we evaluate SED using the multi head memory controlled self attention function in (\ref{eq6}).

\subsection{Model architecture and Training}

\begin{figure}[t]
  \centering
  \includegraphics[width=\linewidth,height=4.5cm,keepaspectratio]{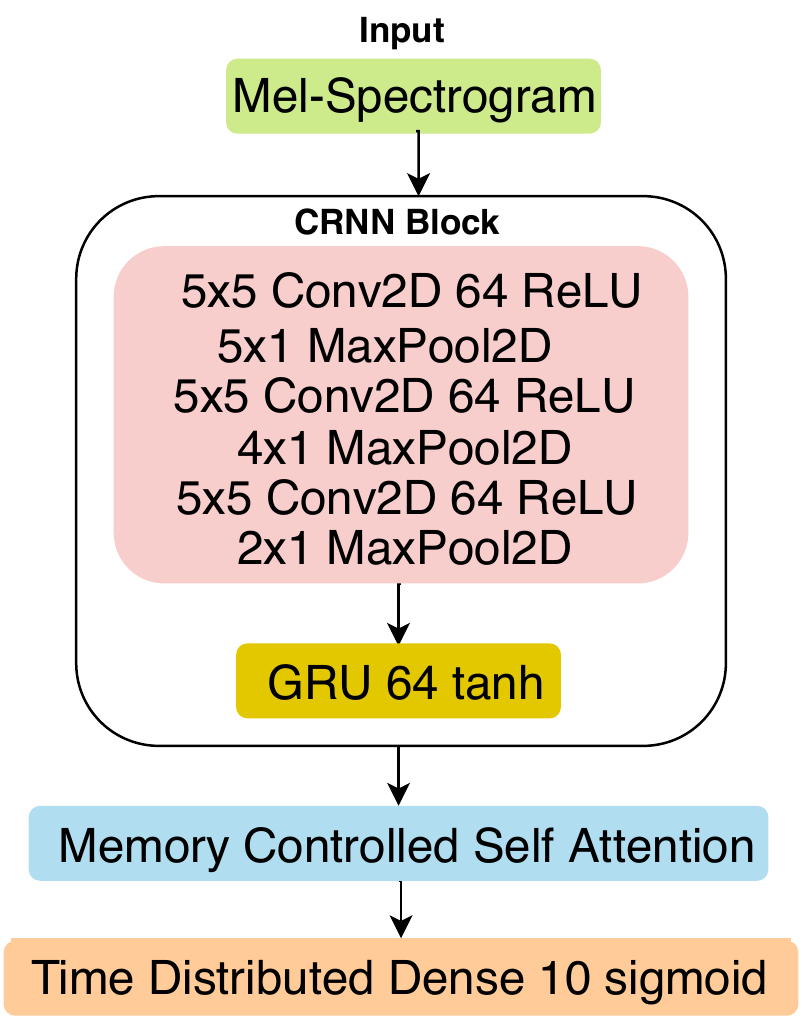}
  \caption{Architecture of SED model.}
  \label{fig:bk}
\end{figure}

We use a similar version of the CRNN model architecture presented in \cite{cakir2017convolutional} to build our SED model; Fig.~\ref{fig:bk} details the models  architecture. We use a $40$ log mel-bands Mel-spectrogram as input representation, extracted using a short-term Fourier transform (STFT) with an FFT window of $2048$, a hop length of $882$, and a sample rate of 44.1 kHz. The CRNN block has three stacked convolutional layers followed by a single gated recurrent unit (GRU) layer. We use a memory controlled self attention layer on the CRNN block feature embeddings. The SED model has a single time distributed dense layer which is the output layer of the network. The output of the model is a posteriogram matrix with dimensions $T\times C$, where $T$ is the number of frames and $C$ is the total number of sound event classes in the dataset. The model predictions are thresholded at $0.5$ to obtain binary two-dimensional representations which are used to compute evaluation metrics based on the ground truth labels.

Each convolutional layer activation is batch normalised and regularised with dropout (probability = $0.3$). The convolutional layer weights have been initialized using random normal distributions with zero mean and $0.05$ standard deviation. We train the network for $200$ epochs using a binary cross-entropy loss function and the Adam optimizer with a learning rate of $0.001$ and a decay of $10^{-6}$. 

\subsection{Dataset and Evaluation metrics}

We train our model on the URBAN-SED \cite{salamon2017scaper} dataset consisting of $10,000$ soundscapes with sound event annotations generated using Scaper \cite{salamon2017scaper}, an open-source library for soundscape synthesis. All recordings are ten seconds long, $16$-bit mono and sampled at $44.1$kHz. The annotations are strong, meaning for every sound event the annotations include the onset, offset, and label of the sound event. Each soundscape contains between one to nine sound events from the list \{air$\_$conditioner, car$\_$horn, children$\_$playing, dog$\_$bark, drilling, engine$\_$idling, gun$\_$shot, jackhammer, siren and street$\_$music\} and has a background of Brownian noise. We use the URBAN-SED pre-sorted train, validation, and test sets. Of $10,000$ soundscapes, $6000$ are used for training, and $2000$ each for validation and test. 

We use the $F$-score and Error Rate (ER), with $F$-score as the primary metric. The evaluation metrics are computed in both segment-wise and event-wise manners using the sed\_eval tool \cite{mesaros2016metrics}. Segment-based metrics show how well the system correctly detects the temporal regions where a sound event is active; with an event-based metric, the metric shows how well the system detects event instances with correct onset and offset. The evaluation scores are micro-averaged values, computed by aggregating intermediate statistics over all test data; each instance has equal influence on the final metric value. We use a segment length of one second to compute segment metrics. The event-based metrics are calculated with respect to event instances by evaluating only onsets with a time collar of $250$ms.

\section{Results and Discussion}

Table~\ref{table:1} presents the SED results. Here, $Baseline$ is the SED model without self attention. $SelfAttn$ is the SED model with self attention and without memory control. $SelfAttn\_L$ is the SED model with memory controlled self attention using attention width $L$. $MultiHead$ is the SED model with memory controlled multi head self attention. 

We see that self attention without memory control has an event-based $F$-score of $9.78\%$ that is significantly lower than the baseline ($20.10\%$) and that the best model ($33.92\%$) uses memory controlled self attention with $L=50$. The model with $L$ = $100$ has an $F$-score of $13.66\%$, which is lower than other memory controlled self attention models. This clearly justifies the need for proper selection of the extent of memory in order to efficiently implement self attention for SED. The inferior performance of the $SelfAttn$ model compared to the $Baseline$ model and the models with memory control is expected and is due to the reasons explained in Section 2. Also, we cannot expect a monotonic model behavior based on the attention width value. The optimum choice of attention width for each audio sample depends on the type of sound events and event durations. The event-based $F$-score for the $MultiHead$ model is $21.89\%$ compared with the best model value of $33.92\%$. We suggest that the soft optimization rule based on the weighted sum of individual attention head representations is the main reason for the under-performance of the $MultiHead$ model. 

\begin{figure}[t]
  \centering
  \includegraphics[width=\linewidth,height=4.6cm,keepaspectratio]{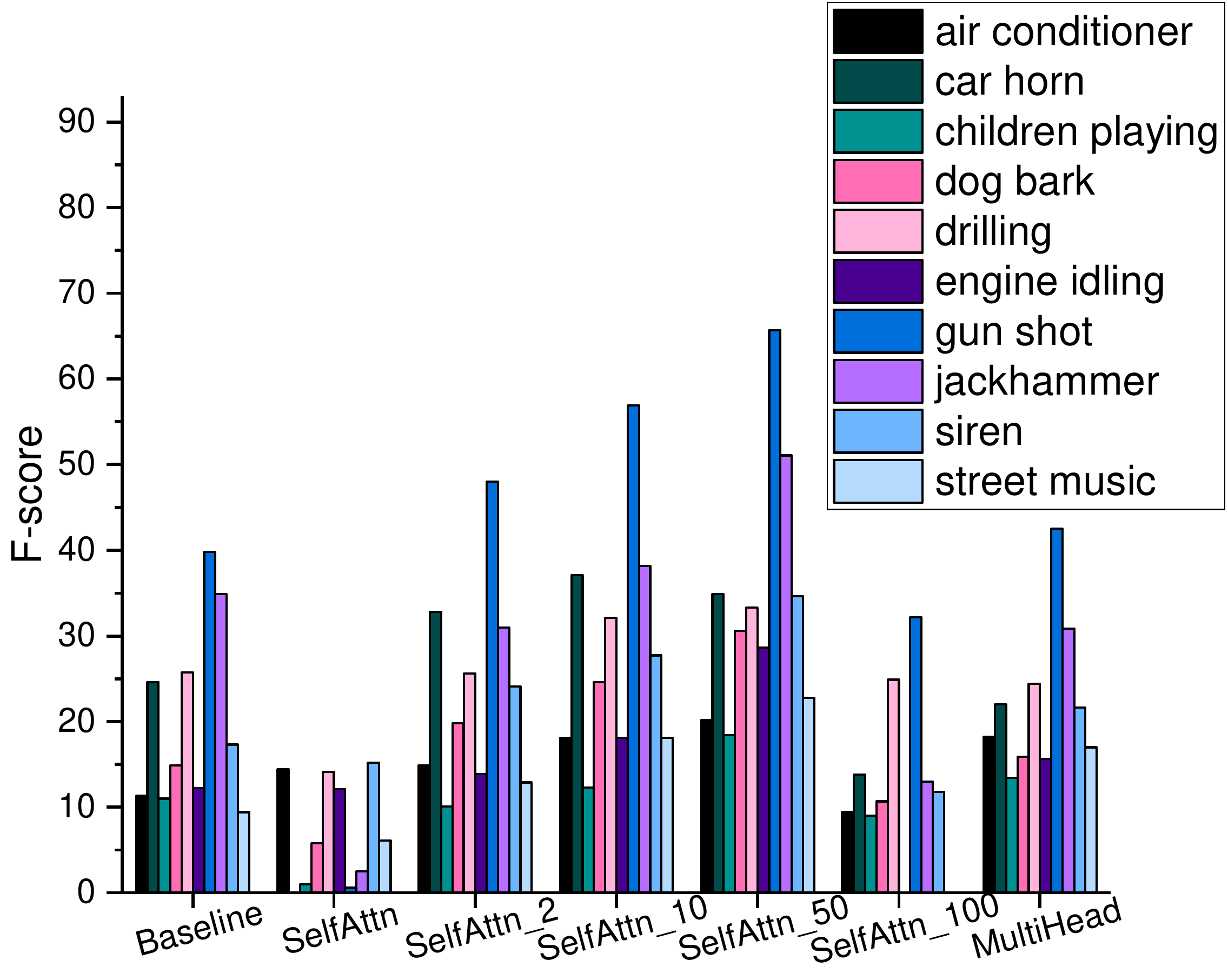}
  \caption{Class-wise event based F-score results.}
  \label{fig:classwise}
\end{figure}

In Fig.~\ref{fig:classwise}, we analyse the class-wise event based $F$-score. We expected best recognition performance for short sound events like $car\_horn$, $dog\_bark$, and $gun\_shot$ with relatively narrow attention width models ($10$ $<$ $L$ $<$ $50$) and for long duration sound events like $drilling$, $engine\_idling$, $air\_conditioner$, and $children\_playing$ using attention models with larger attention width values ($50$ $<$ $L$ $<$ $200$). However for all the sound event classes except $car\_horn$, the memory controlled self attention model with an attention width of 50 frames yields the best recognition performance. As expected, $car\_horn$ being a short event is best recognised with a narrow attention width model ($L$ = $10$). The duration of sound events in the URBAN-SED \cite{salamon2017scaper} dataset is in the range $0.5$ to $4$ seconds. For the effective memory controlled self attention implementation the attention width should not be larger than the duration of short sound events in the dataset. We suggest this is the reason why the attention model with $L$ = $50$ yields the best results. Also, we assume the same reason along with the soft optimization approach for the less effectiveness of the $MultiHead$ model. Even though the overall $F$-score of the $MultiHead$ model is close to the $Baseline$ model, the recognition for long duration events (e.g. air$\_$conditioner, engine$\_$idling) is better with the $MultiHead$ model. Attention visualizations can be found online\footnote{\url{https://github.com/arjunp17/MemoryControlled-MultiheadSelfAtt}}.

\begin{table}[t]
\caption{Sound event detection results.
\label{table:1}}% title of Table
\centering % used for centering table
\begin{tabular}{c c c c c} % centered columns (4 columns)
\hline % inserts single horizontal lines
& \textbf{F1 (\%)} &  & \textbf{Error rate}\\  % inserts table
\hline % inserts single horizontal lines
\textbf{Model} & \textbf{Segment} & \textbf{Event} & \textbf{Segment} & \textbf{Event}  \\ % inserts table
%heading
\hline % inserts single horizontal line
 % inserting body of the table
Baseline & $47.45$ & $20.10$ & $0.74$ & $2.21$ \\
 SelfAttn & $29.45$ & $9.78$ & $0.89$ & $1.51$ \\
 SelfAttn\_$2$ & $50.57$ & $24.44$ & $0.69$ & $1.98$ \\ 
 SelfAttn\_$10$ & $54.36$ & $28.62$ & $0.64$ & $1.68$ \\
 SelfAttn\_$50$ & $\textbf{55.90}$ & $\textbf{33.92}$ & $\textbf{0.59}$ & $\textbf{1.12}$ \\ 
 SelfAttn\_$100$ & $33.71$ & $13.66$ & $0.79$ & $1.21$ \\
 MultiHead & $49.28$ & $21.89$ & $0.69$ & $1.77$ \\% [1ex] adds vertical space
\hline %inserts single line
\end{tabular}
%\label{table:nonlin} % is used to refer this table in the text
\end{table}

\section{Conclusion}

In this work, we investigated the importance of the extent of memory on self attention, applied to the task of sound event detection. Memory controlled self attention is an effective approach to model the relations between frame-level tokens within sound events which improves temporally precise sound recognition. An explicit mapping of the extent of attention to the recurrence relations in audio sequences is a future goal. Our multi-head attention methodology for optimally selecting the extent of attention is not very successful in this work; we are inclined to extend our memory controlled $MultiHead$ model for urban sound tagging using the SONYC \cite{bello2019sonyc} dataset that has a wide range of coarse-grained and fine-grained event tags and also for sound recognition using AudioSet \cite{gemmeke2017audio}. We also see the idea of using memory controlled self attention to define higher level event-based feature embeddings in sound event sequences. 

\bibliographystyle{IEEEtran}

\bibliography{mybib}

% \begin{thebibliography}{9}
% \bibitem[1]{Davis80-COP}
%   S.\ B.\ Davis and P.\ Mermelstein,
%   ``Comparison of parametric representation for monosyllabic word recognition in continuously spoken sentences,''
%   \textit{IEEE Transactions on Acoustics, Speech and Signal Processing}, vol.~28, no.~4, pp.~357--366, 1980.
% \bibitem[2]{Rabiner89-ATO}
%   L.\ R.\ Rabiner,
%   ``A tutorial on hidden Markov models and selected applications in speech recognition,''
%   \textit{Proceedings of the IEEE}, vol.~77, no.~2, pp.~257-286, 1989.
% \bibitem[3]{Hastie09-TEO}
%   T.\ Hastie, R.\ Tibshirani, and J.\ Friedman,
%   \textit{The Elements of Statistical Learning -- Data Mining, Inference, and Prediction}.
%   New York: Springer, 2009.
% \bibitem[4]{YourName17-XXX}
%   F.\ Lastname1, F.\ Lastname2, and F.\ Lastname3,
%   ``Title of your INTERSPEECH 2020 publication,''
%   in \textit{Interspeech 2020 -- 20\textsuperscript{th} Annual Conference of the International Speech Communication Association, September 15-19, Graz, Austria, Proceedings, Proceedings}, 2020, pp.~100--104.
% \end{thebibliography}

\end{document}